\documentclass[palatino,10pt]{article}
\usepackage{moriond,epsfig}

\def\Journal#1#2#3#4{{#1} {\bf #2}, #3 (#4)}

\def\PLB{{\em Phys. Lett.}  B}

\def\EPJC{{\em Eur. Phys. J.} C}

\def\be{\begin{equation}}
\def\ee{\end{equation}}
\def\bea{\begin{eqnarray}}
\def\eea{\end{eqnarray}}

\begin{document}
\vspace*{4cm}
%
%
%
\newcommand{\OPAL}{\textsf{OPAL}}
\newcommand{\ALEPH}{\textsf{ALEPH}}
\newcommand{\DELPHI}{\textsf{DELPHI}}
\newcommand{\LEP}{\textsf{LEP}}
\newcommand{\LEPI}{\textsf{LEP I}}
\newcommand{\LEPII}{\textsf{LEP2}}
\newcommand{\ATLAS}{\textsf{ATLAS}}
\newcommand{\LHC}{\textsf{LHC}}
\newcommand{\LHCB}{\textsf{LHCb}}
\newcommand{\CERN}{\textsf{CERN}}
\newcommand{\SM}{\textsf{SM}}
\newcommand{\MSSM}{\textsf{MSSM}}
\def \lsim{\mathrel{\mathpalette\@versim<}}
\def \gsim{\mathrel{\mathpalette\@versim>}}
\def\SUXU{{SU(2)_L \times U(1)_Y}}
\def\hAp{(\mh,\mA)}
\def\sba2{\sin ^2 (\beta - \alpha)}
\def\cba2{\cos ^2 (\beta - \alpha)}
\def\tanB{\tan \beta}
\def\r{\rightarrow}
\def\Ecm{E_{CM}}
\def\sqs{\sqrt{s}}
\def\h{\mathrm h}
\def\A{\mathrm A}
\def\W{\mathrm W^{\pm}}
\def\Z{\mathrm Z}
\def\HH{\mathrm H^0_{\mathrm{SM}}}
\def\H{\mathrm H}
\def\Hpm{\mathrm H^\pm}
\def\hAA{\h\r\A\A}
\def\ZhA{\Z\r\h\A}
\def\ZshZ{\Zs\r\h\Z}
\def\ZshA{\Zs\r\h\A}
\def\ZsHSMZ{\Zs\r\H\Z}
\def\ZsHZ{\Zs\r\H\Z}
\def\WW {\mathrm W^+ \mathrm W^-} 
\def\ZZ {\mathrm Z \mathrm Z} 
\def\ee{\mathrm e^+\mathrm e^-}
\def\mm{\mu^{+}\mu^{-}}
\def\nn{\nu \bar{\nu}}
\def\qq{\mathrm q \bar{\mathrm q}}
\def\pb{ \mathrm{pb} ^{-1}}
\def\Gcs{\mathrm{GeV/c}^2}
\def\Tcs{\mathrm{TeV/c}^2}
\def\Mcs{\mathrm{MeV/c}^2}
\def\Gc{\mathrm{GeV/c}}
\def\G{\mathrm{GeV}}
\def\eehad{\mathrm e^+\mathrm e^-\rightarrow \rm{hadrons}}
 
\def\Zs{\mathrm Z^{*}}
\def\tt{\tau^{+}\tau^{-}}
\def\ttqq{$\tau^+-\tau^--\mathrm q-\overline{\mathrm q}~$}
\def\ll{\ell^{+}\ell^{-}}
\def\ff{\mathrm f \bar{\mathrm f}}
\def\cc{\mathrm c \bar{\mathrm c}}
\def\bb{\mathrm b \bar{\mathrm b}}
\def\mtau{m_{\tau}}
\def\mmu{m_{\mu}}
\def\mb{m_{\mathrm b}}
\def\mH{m_{\H}}
\def\mHhat{\hat{m_{\H}}}
\def\mh{m_{\h}}
\def\mA{m_{\A}}
\def\mHH{m_{\HH}}
\def\mZ{m_{\Z}}
\def\mW{m_{\mathrm W}}
\def\mt{m_{\mathrm t}}
\def\mS{m_{\mathrm S}}
\def\pt{p_{t}}
\def\G{ \rm{GeV} }
\def\nhit{\rm{N}_{hit}}
\def\mht{\rm{T^{min}_{hemi}}}
\def\ahm{\rm{m^{avg}_{hemi}}}
\def\hnn{\mathrm H^0\nu\overline{\nu}}
\def\ttnn{$\tau^+\tau^-\nu \bar{\nu}$}
\def\hmm{$\mathrm H^0\mu^+\mu^-$}
\newcommand {\gevp}        {${\rm GeV/c }$}
\newcommand{\epsnn}{\mbox{$\epsilon_{\nu \bar{\nu}}$(\%)}}
\newcommand{\Nnn}{\mbox{N$_{\nu \bar{\nu}}$}}
\newcommand{\epsee}{\mbox{$\epsilon_{e^{+}e^{-}}$(\%)}}
\newcommand{\Nee}{\mbox{N$_{e^{+}e^{-}}$}}
\newcommand{\epsmm}{\mbox{$\epsilon_{\mu^{+}\mu^{-}}$(\%)}}
\newcommand{\Nmm}{\mbox{N$_{\mu^{+}\mu^{-}}$}}
\newcommand{\Nexp}{\mbox{N$_{exp}$}}

\title{OBSERVATION OF AN EXCESS IN THE ALEPH SEARCH FOR THE STANDARD 
MODEL HIGGS BOSON}

\author{ Pedro TEIXEIRA-DIAS }

\address{Department of Physics, Royal Holloway University of London, \\
Egham Hill, SURREY TW20 0EX, England}

\maketitle\abstracts{
The ALEPH search for the Standard Model Higgs boson in the year 2000
revealed an excess of signal-like events, consistent with a signal
hypothesis $\mh\approx 115~\Gcs$. Here we present the first results
after the analysis of all of the data collected during the year 2000,
at collision energies up to 209 GeV, and which were published in
November 2000.}

\section{Introduction}

 During the year 2000, the LEP collider was pushed to the edge of its
 performance envelope in order to maximize the Higgs
 boson\,\cite{higgs} discovery potential\,\cite{pj}. The ALEPH
 experiment collected 216.1 $\pb$ of data, at collision energies
 ranging between 202-209 GeV.

 At LEP2, the Higgsstrahlung process, $\ee\r\h\Z$, is the dominant
 mode for producing the Standard Model (SM) Higgs boson. There are
 also smaller contributions from the W- and Z-fusion processes to the
 channels $\ee\r\h\nu_{\mathrm{e}}\bar{\nu_{\mathrm{e}}}$ and $ \h\ee$,
 respectively. 

\begin{figure}
\begin{picture}(400,210)
\put(-15,0){\epsfig{file=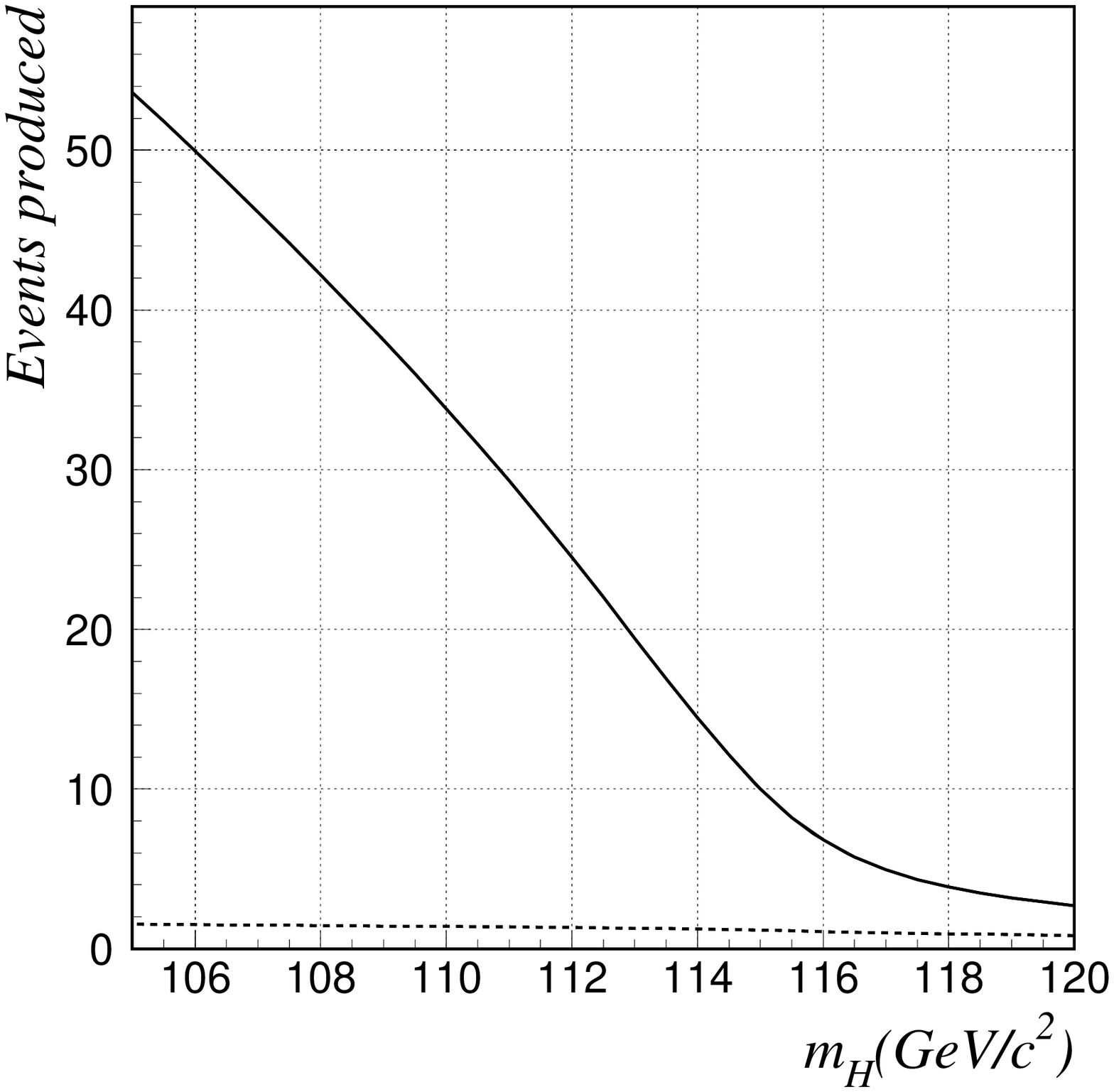,width=8.2cm}}
\put(35,190){(a)}
\put(208,105){\epsfig{file=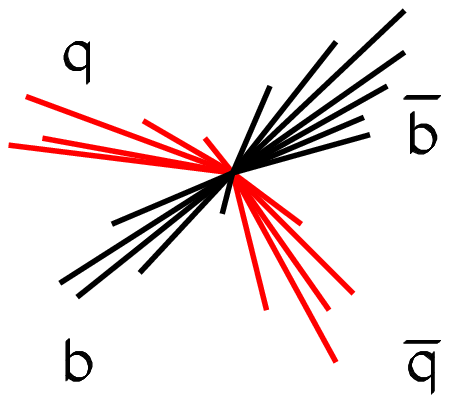,width=3.7cm}}
\put(208,200){(b)}
\put(335,105){\epsfig{file=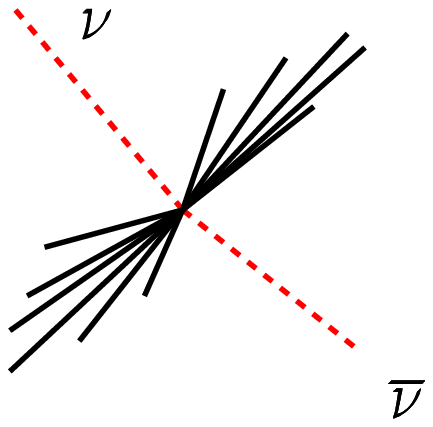,width=3.7cm}}
\put(335,200){(c)}
\put(208,0){\epsfig{file=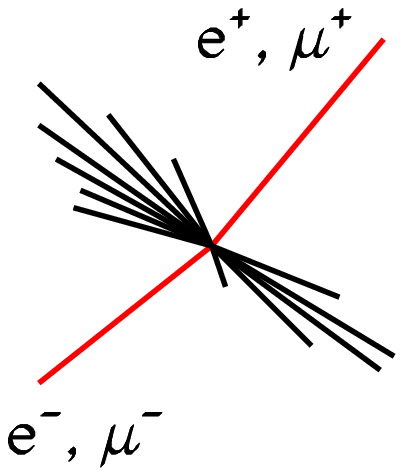,width=3.7cm}}
\put(208,95){(d)}
\put(335,0){\epsfig{file=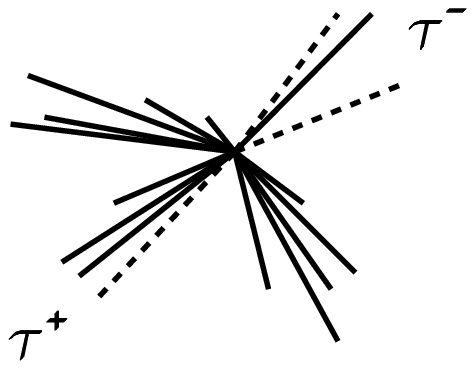,width=3.7cm}}
\put(335,95){(e)}
\end{picture}
\vspace{-0.5cm}
\caption{(a) The expected number of SM Higgs signal events produced 
in the year 2000 data as a function of the Higgs boson mass (solid
curve). The dashed curve is the contribution from the gauge boson
fusion processes, including their interference with the Higgsstrahlung
process. The four main signal event topologies: (b) four-jets
($\h\Z\r\bb\qq$), (c) missing energy ($\h\Z\r\bb\nn$), (d)
leptonic ($\h\Z\r\h\ll$ with $\ell=$e~or~$\mu$) and (e) the final states with
$\tau$ leptons, $\tt\qq$, when either the $\h$ or the $\Z$ decay into a $\tt$
pair.
\label{fig:intro}}
\end{figure}

 Figure \ref{fig:intro}(a) shows the number of Higgs events expected
 to be produced in the data that ALEPH collected, as a function of the
 Higgs boson mass. For an hypothetical Higgs boson mass of 114 (115)
 $\Gcs$ 14.4 (10.0) events are expected. The dominant Higgs decay mode
 for such a signal is $\h\r\bb$ (74\%) followed by $\h\r\tt$
 (7.4\%)~\footnote{For $\mh\sim 115~\Gcs$, the branching fractions to
 $\h\r\WW, gg$ become comparable to $\h\r\tt$, but are not searched
 for explicitly in the ALEPH SM Higgs analysis.}.  Depending on the
 combination of the decay of the Higgs and of the associated Z boson,
 the signal events fall into one of four topologies. These are the
 so-called four-jet, missing energy, leptonic and tau-lepton final
 states, and are described in Figures \ref{fig:intro}(b)-(e).

\section{Higgs search strategy}

 At LEP2 the ALEPH Higgs search strategy rested on two alternative
 analysis ``streams''. While one of the analysis streams relies mostly
 on artificial neural networks (NN) for the event selections, the
 other stream relies mostly on more straight-forward cuts-based event
 selections. The motivation behind this strategy was to provide
 mutually cross-checked results. As the two streams have similar
 performance an eventual signal discovery would have to be confirmed
 in both.

 The two streams differ in the treatment of the two most powerful
 search topologies: four-jets and missing energy. The treatment of the
 $\h\ll$ and $\tt\qq$ channels is in all respects identical between
 the two streams. Table \ref{tab:streams} shows the defining details
 of the cuts-stream and the NN-stream.

\begin{table}[htb]
\footnotesize
\caption{
The two analysis streams: ``cuts'' and ``NN'' denote the type
of event selection used for the given search channel. The observables
$X$ indicate the discriminant variables used for the
calculation of the confidence levels (Section \ref{sec:results}). The
$\h\ll$ and $\tt\qq$ analyses are treated in exactly the same way in the
two streams.\label{tab:streams}}
\vspace{0.4cm}
\begin{center}
\begin{tabular}{|l|c|c|}
\hline
& cuts-stream & NN-stream \\
\hline
\hline
$\h\qq$&  
cuts;$X=m_{\mathrm{rec}}$ &  
NN;$\vec{X}=(m_{\mathrm{rec}}$,NN$_{\mathrm{output}}$)\\
$\h\nn$ &  
cuts;$X=m_{\mathrm{rec}}$ &  
NN;$\vec{X}=(m_{\mathrm{rec}}$,NN$_{\mathrm{output}}$)\\
\hline
$\h\ll$ & 
\multicolumn{2}{|c|}{cuts;$\vec{X}=(m_{\mathrm{rec}}$,b$\tau$-tag)} \\
$\tt\qq$ & 
\multicolumn{2}{|c|}{NN;$X=m_{\mathrm{rec}}$}\\
\hline
\end{tabular}
\end{center}
\end{table}

The actual event selection criteria\,\cite{aleph00} for the different
search channels which were used for the analysis of the 2000 data are
very similar to those used for the 1999 analysis\,\cite{aleph99}. In
particular, the searches used for the four-jet and tau-lepton final
states in 2000 were exactly the same as in 1999. The searches in the
missing energy and leptonic final states were modified
slightly\,\cite{aleph00}. Furthermore, the analysis of the data was
blind, as the event selection criteria were fixed before the start of
the data-taking period.

\section{Search results}\label{sec:results}

Applying the event selections to the 2000 data results in 134 (95)
events being selected in the NN-stream (cuts-stream) while 128.7 (87.6)
are expected from the background. Table
\ref{tab:evsel} shows how the events are distributed between the
four search channels. 

\begin{table}[htb]
\footnotesize
\caption{The number of signal ($s$) and background ($b$) events expected, and 
the number of candidate events ($n_{obs}$) observed in the year 2000
data. For each channel the systematic error on the background is
indicated.  The expected background is divided into $\Z\Z$ (including
$\Z\ee$ and $\Z\nn$), WW (including W$\mathrm{e}\nu$), and $\ff$
(including $\gamma\gamma\r\ff$).  The expectation for the signal 
its significance (Section~3) is computed for a Higgs boson with a mass
of 114\,$\Gcs$.  The numbers from the $\h\ll$ and $\tt\qq$ analyses
are included in both the NN- and cuts-stream totals.
\label{tab:evsel}}
\vspace{0.4cm}
\begin{center}
\begin{tabular}{|c|c|r@{$\pm$}l|r@{$\pm$}l|r@{$\pm$}l|r@{$\pm$}l|c|}
\hline
Analysis & Signal &
\multicolumn{8}{c|}{Background Events} & Events \\
 & Events &
\multicolumn{8}{c|}{Expected} & Obs. \\

\cline{3-10}
 & Expected, $s$  & \multicolumn{2}{c|}{$\Z\Z$} &
\multicolumn{2}{c|}{WW} & \multicolumn{2}{c|}{\rule{0cm}{0.45cm}$\ff$} &
\multicolumn{2}{c|}{Total, $b$} & $n_{obs}$ \\
\hline
\hline
$\h\qq$ (NN) &
4.5  &                     
23.0  & 1.0 &                     
8.6  & 0.6 &                     
15.3  & 1.7 &                     
46.9 & 2.1 &                     
52   \\
\hline
$\h\qq$ (cuts) &
2.9  &                     
12.6 & 0.7 &                     
3.2  & 0.2 &                     
7.9  & 0.7 &                     
23.7 & 1.0 &                     
31   \\
\hline
$\h\nn$ (NN) &
1.4  &                     
13.5 & 0.7 &                     
22.0 & 1.1 &                     
2.0 & 0.4 &                     
37.5 & 1.4 &                     
38  \\
\hline
$\h\nn$ (cuts) &
1.3  &                     
9.9 & 1.1 &                     
8.8 & 1.7 &                     
1.0 & 0.3 &                     
19.7 & 2.0 &                     
20  \\
\hline
$\h\ll$ &
0.7  &                     
26.4 & 0.3 &                     
2.4 & 0.1 &                     
1.8 & 0.3 &                     
30.6 & 0.4 &                     
29   \\
\hline
$\tt\qq$ &
0.4  &                     
6.4 & 0.3 &                     
6.2 & 0.3 &                     
1.0 & 0.3 &                     
13.6 & 0.5 &                     
15  \\
\hline
\hline
NN-stream total &
7.0  &                     
69.3 & 1.3 &                     
39.2 & 1.3 &                     
20.1 & 1.8 &                     
128.7 & 2.6 &                     
134  \\
\hline
Cuts-stream total &
5.3  &                     
55.3 & 1.4 &                     
20.6 & 1.7 &                     
11.7 & 0.9 &                     
87.6 & 2.4 &                     
95 \\
\hline
\end{tabular}
\end{center}
\end{table}

 After the event selection stage, the likelihood of a signal in the
 data is quantified by means of an extended likelihood ratio,
 incorporating information about the numbers of events for both the
 background-only and the signal$+$background hypotheses:

\[
Q = \frac{L_{s+b}}{L_{b}} = \frac{e^{-(s+b)}}{e^{-b}}
\prod_{i=1}^{n_{obs}} \frac{s f_{s}(\vec{X}_{i}) + b f_{b}(\vec{X}_{i})}
{b f_b(\vec{X}_{i})}
\]

 Through $f_s$ and $f_b$, the likelihood ratio $Q$ also contains
 information that allows additional discrimination between the two
 hypotheses. $f_s$ and $f_b$, respectively the signal and background
 probability density functions, are evaluated for each observed
 candidate $i$, with measured discriminant properties $\vec{X}_i$.

 Here, the two analysis streams differ again in the treatment of the
 four-jet and the missing energy searches, in that the cuts-based
 stream uses only the reconstructed Higgs mass as discriminant,
 whereas the NN-based stream uses slightly more powerful
 two-dimensional discriminants (see Table \ref{tab:streams}). Note
 that --in order to keep the selection efficiency high-- the $\h\ll$
 event selection is flavour independent, and that the b/$\tau$ flavour
 of the Higgs candidate jets is only taken into account by means of
 the second discriminant. For this reason, the ALEPH $\h\ll$ search
 for an SM Higgs signal has to retain the second discriminant, even in
 the ``single-discriminant'' cuts-stream.

\begin{figure}
\begin{picture}(400,210)
\put(10,0){\epsfig{file=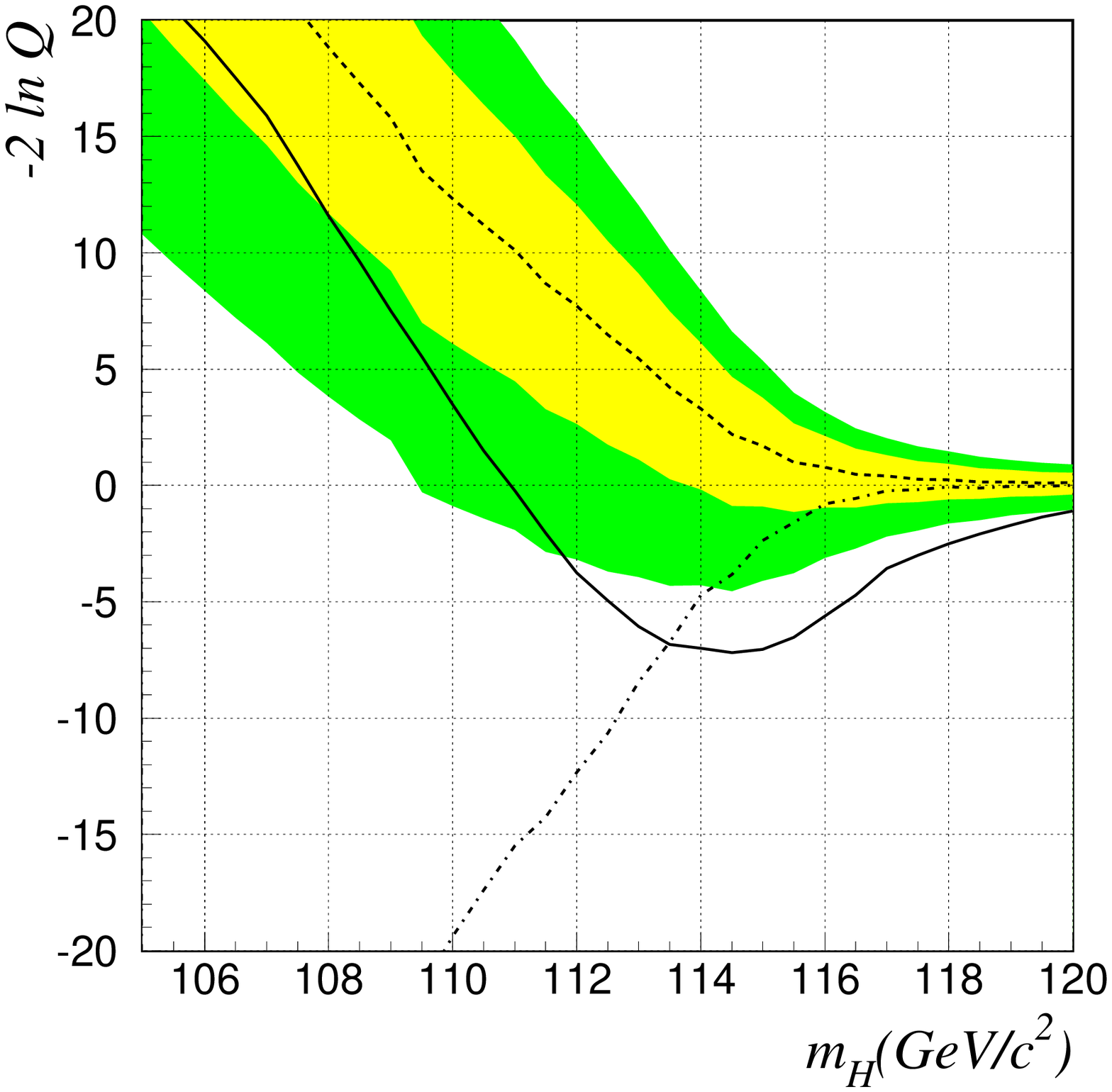,width=7.0cm}}
\put(32,185){(a) {\bf{ALEPH / NN-stream}}}
\put(245,0){\epsfig{file=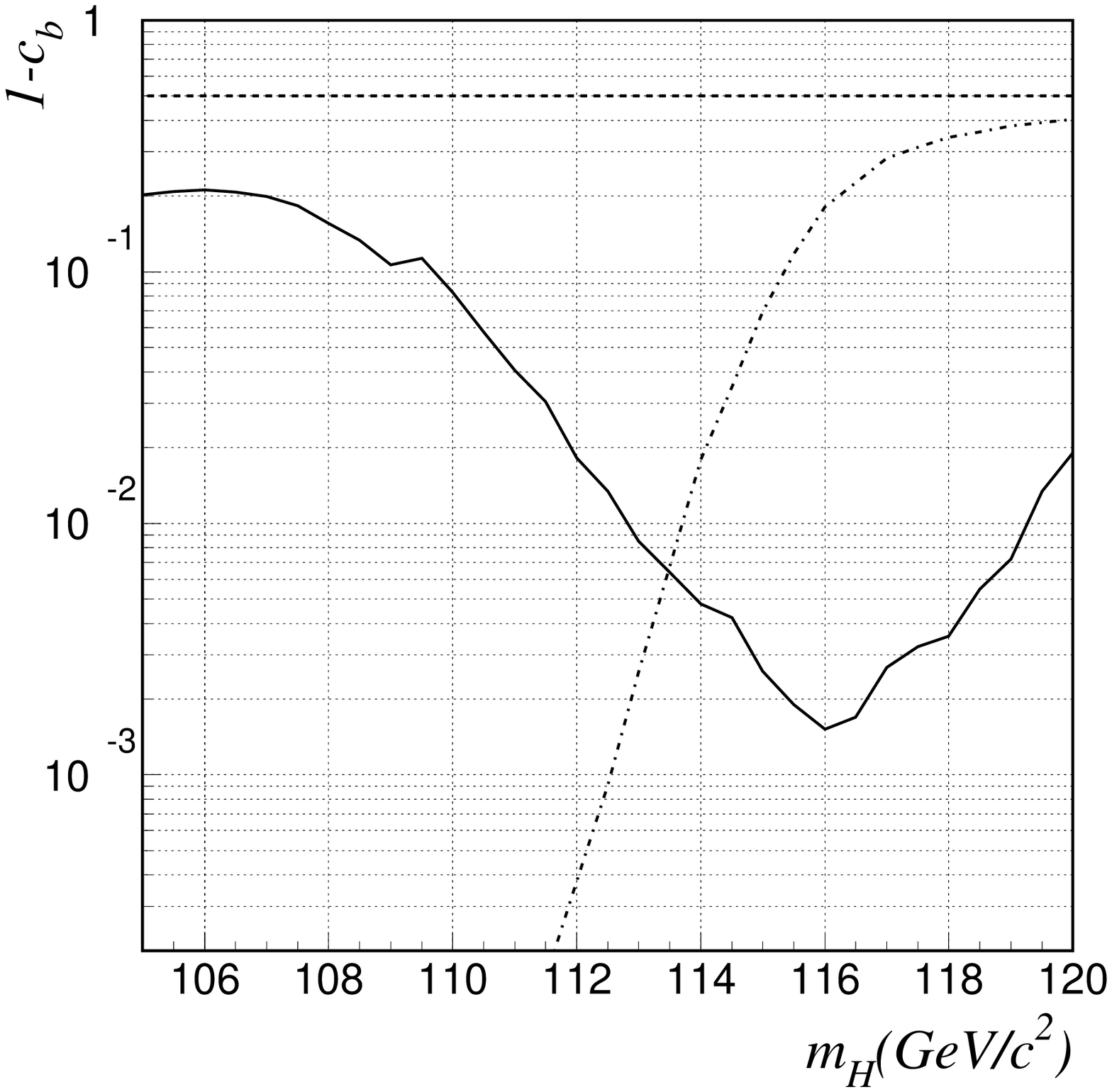,width=7.0cm}}
\put(267,185){(b) {\bf{ALEPH / NN-stream}}}
\end{picture}
\vspace{+0.8cm}
\caption{
(a) The log-likelihood estimator $-2\ln~Q$ as a function of the mass
of the Higgs boson ($\mh$) for the observation (solid) and
background-only expectation (dashed). The light and dark gray bands
around the background-only expectation are the one and two sigma
regions, respectively. The dot-dashed curve shows the expected
position of the median of the log-likelihood estimator when the latter
is calculated at a mass $\mh$ and includes a signal at that same mass.
(b) The observed (solid) and expected (dashed) CL curves for the
background hypothesis as a function of the Higgs boson mass. The
dot-dashed curve indicates the location of the median CL for a Higgs
signal of mass $\mh$.
\label{fig:results_nn}}
\end{figure}

 The likelihood ratio depends on the mass, $\mh$, of the signal
 hypothesis being tested. Figure \ref{fig:results_nn}(a) shows the
 value of $-2\ln~Q$ for the NN-stream, as a function of $\mh$. The
 minimum of this curve indicates a high likelihood for the signal
 hypothesis $\mh\approx~115~\Gcs$. At this mass, the likelihood of the
 signal+background hypothesis is $\sim~$34 times larger than the
 background-only likelihood.

As can be seen from inspecting the grey bands around the
background-only expectation curve (Figure \ref{fig:results_nn}(a)), it
is not impossible (but rather unlikely) that the observed result at
$\mh\approx~115~\Gcs$ could be due to a (large) fluctuation of the
background. In order to quantify the probability of such a
fluctuation, the fraction of background-only Gedanken experiments that
are at least as signal-like as observed is calculated. This fraction,
$1-c_b$ (where $c_b$ is the confidence level on the background-only
hypothesis), is shown in Figure
\ref{fig:results_nn}(b). 
For the background only scenario $1-c_b$ has a median value of 0.5,
whereas if the data contains a signal $1-c_b$ is expected to have a
localized dip to lower values. The minimum of the curve, at $1.5\times
10^{-3}$ probability, corresponds to a $3.0\sigma$ excess over the
expected background.

\section{Discussion and Cross-checks} 

The results presented here have been extensively cross-checked, with a
view to find any possible systematic effects that could significantly
affect the main conclusions of the searches. These checks are
summarily described here. The reader is referred to the
publication\,\cite{aleph00} for more details.

The differences between the cuts-stream and the NN-stream can be
summarized as:
\begin{itemize}

\item In the cuts-stream (NN-stream) 3 out of the 4 event selections are 
based on cuts (neural networks);
\item In the cuts-stream 3 out of the 4 discriminants are one-dimensional 
($m_{\mathrm{rec}}$) whereas in the NN-stream 3 out of the 4
discriminants are two-dimensional (an additional discriminant is used
in addition to $m_{\mathrm{rec}}$).
\end{itemize}

 Table \ref{tab:compare} compares likelihood ratio and significance
 values between the two streams. It can be readily seen that the
 results from the two analysis streams are in agreement. (The complete
 curves for the cuts-stream can be found elsewhere\,\cite{aleph00}.)
 For the cuts-stream $1-c_b$ has a minimum of $1.1\times 10^{-3}$,
 corresponding to a $3.1\sigma$ excess with respect to the expected
 background.

\begin{table}[htb]
\footnotesize
\caption{
Comparison of the main findings in the cuts-stream and the
NN-stream. The Higgs boson masses are indicated in $\Gcs$. The most
significant candidates in each stream are indicated (see Table
\ref{tab:4jcands} for details).
(The LEP Higgs working group uses a different convention to compute
these significances. Using that double-sided Gaussian convention, the
significances quoted here would be increased by
$\approx0.2\sigma$.)\label{tab:compare}}
\vspace{0.4cm}
\begin{center}
\begin{tabular}{|l|cccc|}
\hline
            & $-2\ln~Q$   & significance & maximum      & significant\\
            & ($\mh$=115) & ($\mh$=115)  & significance & candidates \\
\hline
\hline
cuts-stream & -5.7 & $2.9\sigma$ & $3.1\sigma$ & $b,c,d,e$ \\
\hline
NN-stream   & -7.0 & $2.8\sigma$ & $3.0\sigma$ & $a,b,c$   \\
\hline
\end{tabular}
\end{center}
\end{table}

\begin{table}[htb]
\footnotesize
\caption{
Details of the five four-jet candidates selected with an event weight
\(\ln(1+\frac{sf_s}{bf_b})\) 
greater than 0.5 at a Higgs boson mass of 115$~\Gcs$ in either the NN
or cut streams. The jet b-tag values vary from 0 (non-b-like) to 1
(b-like). The four-jet NN output approaches 0 (1) for background-like
(signal-like) events.\label{tab:4jcands}}
\vspace{0.4cm}
\begin{center}
\begin{tabular}{|c|c|c|c|c|c|c|c|}
\hline
Candidate & $\sqrt{s}$ & $m_{\mathrm{rec}}$ & 
\multicolumn{4}{c|}{b-tagging} & 4-jet \\
\cline{4-7}
(Run/Event) & (GeV) & ($\Gcs$)& jet 1 & jet 2 & jet 3 & jet 4 & NN \\
\hline
\hline
$a$ (56698/7455) & 206.5 & 110.0 & 0.999 & 0.836 & 0.999 & 0.214 & 0.999
 \\		   	           
\hline                                       
$b$ (56065/3253) & 206.7 & 112.9 & 0.994 & 0.776 & 0.993 & 0.999 & 0.997
 \\		   	           
\hline                                       
$c$ (54698/4881) & 206.7 & 114.3 & 0.136 & 0.012 & 0.999 & 0.999 & 0.996
 \\		   	           
\hline                                       
$d$ (56366/0955) & 206.5 & 114.5 & 0.238 & 0.052 & 0.998 & 0.948 & 0.935
 \\		   	           
\hline                                       
$e$ (55982/6125) & 206.7 & 114.6 & 0.088 & 0.293 & 0.895 & 0.998 & 0.820
 \\
\hline                                                             
\end{tabular}
\end{center}
\end{table}

 Most of the observed effect originates in the four-jet channel: two
 $\bb\bb$ events ($a, b$) and three $\bb\qq$ events ($c, d, e$). Table
 \ref{tab:4jcands} gives details of these events.  

 The similarity of the findings in the two streams, in itself dispels
 any concerns about the results being a construct of the NN selection
 or of the more sophisticated two-dimensional discriminant. The
 comparison of the two streams does not however allow any inferences
 regarding common points between the four-jet analyses, such as the
 reconstructed mass discriminant and (to some extent) the choice of
 the pairing.

 In the four-jet channel $(E,p)$ conservation is imposed by means of a
 4C-kinematic fit. The reconstructed mass \( m_{\mathrm{rec}} = m_{12}
 + m_{34} - m_Z\) is calculated using the fitted masses for the Z and
 Higgs candidate dijets in the event, $m_{12}$ and $m_{34}$
 respectively. No evidence of a bias towards the hZ threshold ($\sim
 \sqrt{s}-m_Z$) has been found either in the 1999 data ($\sqrt{s}=$192
 - 202 GeV) or in the 2000 data with $\sqrt{s} < $206 GeV. In
 addition, the reliability of the mass reconstruction was confirmed by
 applying it to a large control sample of four-jet WW events selected
 with the cuts analysis, which had been slightly modified to include
 an anti-b-tagging cut.

 The choice of pairing in the selected four-jet events (i.e., choosing
 which of the dijets corresponds to the Higgs candidate and which to
 the Z candidate) can obviously affect the reconstructed mass spectra.
 Using the cuts analysis, it has been shown\,\cite{aleph98} that the
 decay angles of the h and Z candidate dijets can be successfully used
 to select the best pairing and thus improve discrimination with
 respect to the background. The decay angles are especially useful in
 the case of $\bb\bb$ events, where the b-tagging information is of no
 value for the pairing choice. The decay angles were therefore also
 incorporated in the four-jet NN~\footnote{For events selected by the
 NN, the chosen pairing is that with the largest NN output. In
 practice this means that the NN pairing choice is essentially based
 on the values of $m_{12}$, the jet b-tags and the decay angles.}. The
 choice of pairing using the decay angles is well modelled in the
 simulation and has been used for the large data samples collected at
 $\sqrt{s}=192-202$ GeV (and, for the cuts-analysis, at $\sqrt{s}=189$
 GeV as well) without any evidence of a bias.

 For additional systematic checks the reader is referred
 elsewhere\,\cite{aleph00}.

\section{Conclusion}
The ALEPH search for the SM Higgs boson resulted in the observation of
an excess of signal-like events in collisions with $\sqrt{s}>$ 206
GeV, consistent with the $\mh\approx\,115\,\Gcs$ signal hypothesis. This
excess, with {\sl{ca.}}  3.0$\sigma$ significance, has been observed
in the two ALEPH alternative searches: the cuts-stream and the
NN-stream. A final publication, including more detailed systematic
studies, is in preparation.

 The other LEP collaborations have also published their search
 results\,\cite{dlo}. DELPHI and OPAL have presented their results at
 this conference\,\cite{thesep}. The combined result of the SM Higgs
 boson search at LEP has also been presented here\,\cite{myproc2}.


\section*{References}
\begin{thebibliography}{99}

\bibitem{higgs}
P.W. Higgs, Phys. Lett. {\bf 12} (1964) 132;
Phys. Rev. Lett. {\bf 13} (1964) 508; Phys. Rev. {\bf 145} (1966) 1156;\\
F. Englert and R. Brout, Phys. Rev. Lett. {\bf 13} (1964) 321;\\
G.S. Guralnik, C.R. Hagen, and T.W.B. Kibble, Phys. Rev. Lett.
{\bf 13} (1964) 585;\\
T.W.B. Kibble, Phys. Rev. {\bf 155} (1967) 1554.

\bibitem{pj} 
P. Janot {\sl{Priorities for LEP in 2000}} 
in {\em Proceedings of the 10th Workshop on LEP-SPS performance}, 
P. Le Roux, J. Poole, M. Truchet (Eds.); CERN-SL/2000-007.

\bibitem{aleph00} 
The ALEPH Coll., R. Barate {\it et al.}, \Journal{\PLB}{495}{1-17}{2000}.

\bibitem{aleph99} 
The ALEPH Coll., R. Barate {\it et al.}, \Journal{\PLB}{499}{53-66}{2001}.

\bibitem{aleph98}
The ALEPH Coll., R . Barate {\it et al.}, \Journal{\EPJC}{17}{223-240}{2000}.

\bibitem{dlo}
The DELPHI Coll., P.Abreu {\it et al.}, \Journal{\PLB}{499}{23-37}{2001};\\
The L3 Coll., M. Acciarri {\it et al.}, \Journal{\PLB}{495}{18-25}{2001};\\
The OPAL Coll., G. Abbiendi {\it et al.}, \Journal{\PLB}{499}{38-52}{2001}.

\bibitem{thesep}
I. Nakamura (OPAL) and P. Morettini (DELPHI), these proceedings.

\bibitem{myproc2}
P. Teixeira-Dias, on behalf of the LEP Higgs Working Group, {\sl{The
SM Higgs boson search at LEP: combined results}}, these proceedings.

\end{thebibliography}
\end{document}